\documentclass[12pt,a4paper]{article}

\def\SF{S_F^2}
\def\LP{l_{P}}
\def\MP{{m_{P}}}
\def\T{{\bf T}}
\def\X{{\bf X}}
\def\Z{{\bf Z}}
\def\bfL{{\bf L}}
\def\beq{\begin{equation}}
\def\eeq{\end{equation}}
\begin{document}

\title{
\hfill \hbox{\small DIAS-STP-04-12}\\
Quantum Black Holes: \\
the Event Horizon as a Fuzzy Sphere\\  }
\author{Brian P. Dolan\footnote{\tt bdolan@thphys.may.ie}\\
{\sl Dept. of Mathematical Physics, NUI, Maynooth, Ireland}\\
and \\
{\sl School of Theoretical Physics}\\
{\sl Dublin Institute for Advanced Studies}\\ 
{\sl 10~Burlington Rd., Dublin 8, Ireland}\\
\\}

\maketitle

\begin{abstract}

Modeling the event horizon of a black hole
by a fuzzy sphere it is shown that 
in the classical limit, for large astrophysical black-holes,
the event horizon looks locally like
a non-commutative plane with non-commutative parameter dictated
by the Planck length.
Some suggestions in the literature concerning
black hole mass spectra are used to
derive a formula for the mass spectrum of
quantum black holes in terms of four integers which
define the area, angular momentum, electric and magnetic
charge of the black hole.  
We also suggest how the classical bounds on extremal black holes
might be modified in the quantum theory.

\end{abstract}

\section{Introduction} 

Bekenstein's suggestion 
that the surface area of a black hole is related to entropy and 
that the entropy should in fact be
proportional to the area \cite{Bekenstein}, was triumphantly
vindicated by Hawking's calculation of the black hole temperature 
and entropy as a function of area \cite{Hawking}. 
If the entropy is to be finite it then necessary that there be
a finite number of degrees of freedom associated with
the event horizon area -- it should be quantised \cite{Sorkin}.
Quantising the event horizon is very reminiscent
of the concept of a \lq\lq fuzzy sphere'', $S_F^2$ \cite{Madore},
in which points are \lq\lq smeared out'' and the
geometry becomes non-local.
In this paper we shall investigate modeling a black hole event
horizon with a fuzzy sphere and show that this idea fits nicely
with many of Bekenstein's suggestions of treating
a black hole as a particle, \cite{Particle} \cite{Area}
(a point of view also strongly advocated by 't~Hooft \cite{tHooft2}).

It has been suggested that the area
of a black hole should have a quantised spectrum
\beq A=a (N+\eta)\LP^2,
\label{BekensteinAreaLaw}\eeq
with $N=1,2\ldots$, and $a>0$, $\eta>-1$ 
undetermined constants ($\LP=\sqrt{G_N\hbar/c^3}$
is the Planck length), \cite{Sorkin,Area,tHooftBombelli}).
This idea has since been developed further in 
\cite{OtherArea} and discretisation of the horizon has 
also been postulated by 't Hooft \cite{tHooft1}.  
It was suggested some time ago 
that a black hole event horizon might be modeled by a fuzzy sphere
\cite{MadoreBook}.

It is shown in section 2 that, in a fuzzy sphere model in the classical
limit $N\rightarrow \infty$, the neighbourhood of a point
on the event horizon locally looks like 
a non-commutative plane with non-commutativity
parameter
\beq\theta={a\LP^2\over 4\pi}\eeq
where $a$ is a numerical constant of order one related to the event 
horizon area by (\ref{BekensteinAreaLaw}).
A relation between quantisation of the event horizon area
and the non-commutative plane was suggested in \cite{Romero}.
Non-commutativity on the event horizon was also suggested
in \cite{Li} and
a direct approach to deriving non-commutativity in black hole
physics was recently initiated in \cite{BuricMadore}.

Part of the characterisation of a fuzzy sphere is
an irreducible representation of $SU(2)$ of dimension
$N=2k+1$, with $k$ either integral or half-integral.
Functions on the fuzzy sphere are then represented
by $N\times N$ matrices acting on an $N$-dimensional
Hilbert space. 
We argue in the following that
it is natural to take the area of the event horizon to
be 
\beq A=4\pi(2k+1)\LP^2
\label{FourPiAreaLaw}
\eeq
so that $a=4\pi$ and $\eta=0$ above.   
The value of $a=4\pi$ that is natural in a fuzzy sphere construction
has also been found in the semi-classical
approach of \cite{PadPat} and a
mini-superspace approach to black hole
quantisation in \cite{Gabor} \cite{Medved}.
An equal spaced area spectrum like that of (\ref{FourPiAreaLaw}) 
was found in
\cite{Polychronakos} though the prefactor was undetermined,

With the values $a=4\pi$ and $\eta=0$ above
we show that the mass spectrum 
for black holes suggested by Bekenstein \cite{Area}
is modified to give:
\begin{equation}
M_{k,j,q_e}^2
=\left\{ {\left(2k+ 1 +\alpha q_e^2 \right)^2
+4j(j+1) \over 4(2k+1)} \right\}\MP^2,
\end{equation}
where $j$ is integral or half-integral and $q_e$ is an integer,
representing angular momentum
and electric charge respectively, $\alpha=e^2/\hbar c$ is the
fine structure constant and $\MP=\sqrt{c\hbar/ G_N}$ is the Planck mass
(there is a modification
of this formula when magnetic monopoles are included).
The smallest possible mass for a black hole in this scheme is therefore
\beq M={1\over 2}m_P,\eeq
when $k=j=q_e=0$.

For given $j$ and $q_e$ the quantum number $k$ is bounded below
by
\beq
(2k+1)^2\ge 4j(j+1) + \alpha^2 q_e^4.
\eeq
In particular, for a zero charge black hole, the
classical bound
\beq
J^2\le M^4\eeq
(in units with $G_N=c=1$) is replaced by
\beq
J^2\le M^4 -{\pi^2\LP^4\over A^2}\hbar^2.\eeq

The layout of the paper is as follows.
In section \ref{BlackHoles} the quantisation of the area arising
from the fuzzy sphere hypothesis is discussed for Schwarzschild
black holes and the projection to the non-commutative plane is
explained.  Section \ref{RotatingHoles} analyses non-zero angular
momentum and the associated bounds on the mass while
section \ref{ChargedHoles} does the same for charged and rotating holes. 
The relation to entropy is discussed in section \ref{EntropySection}
and the results are summarised in section \ref{Conclusions}

\section{Schwarzschild Black Holes}
\label{BlackHoles}

The 2-dimensional sphere is a symplectic manifold --- a phase-space
in physics language, albeit a compact one.
This phase-space can be quantised to give $\SF$.
The concept of a point on $\SF$
is not defined
but instead the points are smeared out into a finite number
of phase-space
\lq cells', hence the name \lq fuzzy', \cite{Madore}.
For any integer, $N=2k+1$ with $k$ labelling $SU(2)$ 
representations either integral or half-integral,
$\SF$  has $N$ cells and operators
on phase-space are $N\times N$ matrices acting on a $N$-dimensional
Hilbert space, \cite{Perelomov}. Visually $\SF$ might be viewed
as being 
like the surface of Jupiter, with the
belts being unit cells, but this is not essential since, as in
any quantum phase space,
only the area of the fundamental cells, not their shape, is fixed.

If we picture the event horizon of a black hole as a fuzzy sphere
then the total area of the event horizon is naturally a multiple
of the area of a fundamental unit cell.
Suppose the unit cells have area $a\LP^2$, with $a$ a positive
dimensionless constant of order one.  Then the total area 
of the event horizon is
\begin{equation}
A=Na\LP^2,
\label{AreaN}
\end{equation}
and, since $N=2k+1$, we conclude that $\eta=0$ in equation 
(\ref{BekensteinAreaLaw}).

For a non-rotating black hole with zero charge (\ref{AreaN}) immediately
implies that 
the Schwarzschild radius $R_S$ is also quantised
\beq
R^2_S=A/4\pi={Na\LP^2\over 4\pi}.\eeq
To avoid messy factors of $4\pi$ it is  
convenient to define $\bar A=A/4\pi$
and $\bar a = a/4\pi$
so 
\beq
R^2_S=\bar A=N\bar a \LP^2.\eeq
The mass of the hole can now be expressed as 
\beq
M={R_S c^2\over 2G_N}=\sqrt{N \bar a}{\LP c^2 \over 2G_N}=
\sqrt{N\bar a} {\MP\over 2}.\eeq
The hypothesis that the event horizon is a fuzzy-sphere
thus immediately leads us to conclude that black hole
masses are quantised 
\begin{equation}
M^2={N\bar a\over 4}\MP^2
\label{quantisedmass}
\end{equation}
with $N$ a positive integer.

For astrophysical black holes $N$ is so large that the quantum
nature of the mass would be unobservable, but in the final stages of
black hole evaporation the black hole would go through a series
of discrete states until the final state is reached, with $N=1$
({\it i.e.} $k=0$)
and residual mass $M_0=\sqrt{\bar a}\MP/2$.  Thus in this picture
evaporating black holes do not disappear but must necessarily
leave behind a residual hole of the order of the Planck mass.
As remarked in \cite{Particle} 
the situation is reminiscent of the Bohr model of the
atom in which orbiting electrons can only occupy a discrete
set of orbits, dictated by the Bohr-Sommerfeld constraint 
$\oint pdq=2\pi N\hbar$ on the orbitals,
and decaying electrons must finally lodge in the ground state
thus rendering atoms stable.

Non-commuting co-ordinates on the fuzzy-sphere can
be represented globally by three $N\times N$ matrices $\X_i$,
$i=1,2,3$, satisfying
\beq
\X_i\X_i=R_S^2{\bf 1},\eeq
where ${\bf 1}$ is the $N\times N$ unit matrix, with $\X_i$ 
proportional
to the generators $\bfL_i$ of $SU(2)$ in the irreducible $N\times N$
representation,
\beq
[\bfL_i,\bfL_j]=i\epsilon_{ijk}\bfL_k, \qquad 
\bfL_i\bfL_i=k(k+1){\bf 1}.\eeq
From this we deduce that
\begin{equation}
\X_i=\lambda_k\bfL_i\qquad
\Rightarrow \qquad [\X_i,\X_j]=i\lambda_k\epsilon_{ijk}\X_k,
\label{Rsj}
\end{equation}
with 
\beq
\lambda_k:=
\sqrt{\bar a}\LP\sqrt{2k+1\over k(k+1)}.\eeq

At first glance it appears that, in the large $N$ limit, 
the $\X_i$ in equation
(\ref{Rsj}) become commutative and the commutative sphere
is recovered, since $\lambda_k\rightarrow 0$ in the limit,
but upon more careful consideration this is not 
in fact correct.\footnote{An earlier version of this paper 
contained an error on
this point and  I am grateful to Al Stern and Eli Hawkins for bringing this to my attention.}
Heuristically this can be seen by focusing on a region near the
south pole of a large black-hole, in the limit of large $k$.
At the south pole $\X_1$ and $\X_2$ are transverse to the surface
and $\X_3$ is normal to it, with $\X_3\approx -R_s$ and
\beq
[\X_1,\X_2]=i \lambda_k \X_3 = i \lambda^2_k {\bf L}_3.  
\eeq     
In a basis in which
\beq
{\bf L}_3=\pmatrix{k & & \cr & \ddots & \cr & & -k\cr}
\eeq
is diagonal the eigenvalue $\X_3\approx -R_s$ corresponds to the 
minimum eigenvalue
$-k$ of ${\bf L}_3$ so
\beq
[\X_1,\X_2]\approx -i\lambda_k^2 k 
\eeq
and, as $k\rightarrow\infty$,
\beq
[\X_1,\X_2]=-2i\bar a l_P^2.
\eeq
Hence, in an infinitesimal region around the pole, the
event horizon looks like a non-commutative plane in the infinite $k$
limit.

This observation can be put on a more formal footing using the
analysis of 
\cite{Al} (see also \cite{Chuetal}) in which is shown that the
$k\rightarrow\infty$ limit of (\ref{Rsj}) describes a non-commutative
plane under stereographic projection.  
This is seen by defining $\X_\pm=\X_1\pm i \X_2$
and performing the analogue of stereographic projection for
fuzzy co-ordinates:
\beq
\Z =\X_- ({\bf 1} - \X_3/R_S)^{-1}, 
\qquad \Z^\dagger = ({\bf 1}-\X_3/R_S)^{-1}\X_+.\eeq
Then, for large $k$, 
\beq
[\Z,\Z^\dagger]=2\lambda_k R_S({\bf 1}-\X_3/R_S)^{-2}+ o(1/k). 
\label{Zcommutator}
\eeq
Now, although the operator $\X_3/R_S$ has eigenvalues between $-1$ to $+1$
inclusive,
only a very small range above $-1$ is necessary to cover the whole
$\Z$-plane.
To see this observe that
\beq
{1\over 2}(\Z\Z^\dagger + \Z^\dagger \Z)=R_S^2\left( {\bf 1}+
{\X_3\over R_S}\right)
\left( {\bf 1}-{\X_3\over R_S} \right)^{-1} + o(1/k).
\eeq
Writing $\X_3/R_S=-{\bf 1}+\T/R_S^2$, where $\T/R_S^2$ has eigenvalues
between $0$ and $2$,
this reads
\beq
{1\over 2}(\Z\Z^\dagger + \Z^\dagger \Z)={1\over 2}\T\left({\bf 1}-{\T\over 2R_S^2}\right)^{-1}
+ o(1/k).
\label{Zplane}
\eeq
Now, in the $k\rightarrow\infty$ limit,
we can cover the whole of the $\Z$-plane by projecting
all operators 
onto the subspace spanned by of eigenvectors of $\T$ 
with eigenvalues in the
range $0$ to $\bar a \sqrt{k}\,\LP^2$.
Hence, for $k\rightarrow\infty$, $\T/R_S^2\rightarrow 0$ 
in (\ref{Zplane}) and we can replace $\X_3\over R_S$ with 
$-{\bf 1}$ in (\ref{Zcommutator}) to give
\beq
\Z\Z^\dagger + \Z^\dagger \Z =\T \qquad\hbox{and} \qquad
[\Z,\Z^\dagger]=\theta  \eeq
with non-commutativity parameter
\beq
\theta = \lim_{k\rightarrow\infty}{\lambda_k R_S\over 2} 
= \lim_{k\rightarrow\infty}{\bar a\LP^2
\over 2}{(2k+1)\over \sqrt{k(k+1)}}
={\bar a\LP^2}.\eeq
The interesting conclusion of this analysis is that, even for
large astrophysical black-holes, there is a vestige of non-commutativity
at the Planck length.  If the assumptions made here are correct
the event horizon of a black-hole is
a physical example of a system in which Connes' non-commutative
geometry manifests itself in the continuum.

\section{Rotating Black Holes}
\label{RotatingHoles}

Now consider a rotating black hole with angular momentum
$J^2=j(j+1)\hbar^2$ 
and zero charge. The event horizon is still topologically 
a sphere, though not metrically a round sphere
it still has a fuzzy description.
The classical formula for the mass as a function of
angular momentum and area (the Christodoulou-Ruffini 
mass \cite{Christodoulou}) is
\footnote{Here we use units in which $G_N=c^2=1$ to keep the formula
clean, but $\hbar$ will be retained
so as to highlight quantum phenomena.  
Hence $\LP^2=\MP^2=\hbar$.}

\begin{equation}
M^2={1\over 4} \bar A + {J^2\over \bar A},
\label{CRmass}
\end{equation}
or
\beq
{\bar A\over 2}=M^2+\sqrt{M^4-J^2}
\label{Jarea}
\eeq
(the positive square root is taken here because $A$ is the area
of the outer horizon).
From the above formula comes the bound 
\begin{equation}
J^2\le M^4,
\label{Jbound}
\end{equation}
otherwise $\bar A$ becomes complex.  Using (\ref{CRmass}) and
(\ref{Jarea}) this is equivalent to
\beq
J^2\le {1\over 4}\bar A^2.
\eeq

Classically the maximum allowed angular momentum is when
(\ref{Jbound}) is saturated: 
\begin{equation}
J_{max}^2=M^4={1\over 4} \bar A^2.
\label{Jmax}
\end{equation}

Consider the quantum version of (\ref{Jmax}).
Using $J^2_{max}=j_{max}(j_{max}+1)\hbar^2$,
together with 
the ansatz (\ref{AreaN}),
gives
\begin{equation}
\left(j_{max}+{1\over 2}\right)^2=
\bar a^2\left(k+{1\over 2}\right)^2+{1\over 4}.
\end{equation}
Quantum mechanically the bound might not be saturated so 
all we can safely say is that
\begin{equation}
\left(j_{max}+{1\over 2}\right)^2\le
\bar a^2\left(k+{1\over 2}\right)^2+{1\over 4}.
\label{QJMax}
\end{equation}

Suppose that the bound is saturated in the limit of large $k$,
and hence large $j_{max}$, so that
\beq\lim_{k\rightarrow\infty}{J^2_{max}\over M^4}=1\quad
\Leftrightarrow \quad
\lim_{k\rightarrow\infty}{4 J^2_{max}\over \bar A^2}=1\quad
\Leftrightarrow \quad
\lim_{k\rightarrow\infty}{j^2_{max}\over k^2}=\bar a^2.\eeq

Now the fuzzy sphere is associated with a Hilbert
space whose maximum angular momentum is $k$, so it seems
very natural to take $j_{max}=k$, in which case $\bar a=1$.
Then (\ref{Jmax}) must be modified to read
\begin{equation}
J_{max}^2/\hbar^2=j_{max}(j_{max}+1)={1\over 4}(\bar A^2/\hbar^2-1)
\label{QuantumJmax}
\end{equation}
with
\beq
 \bar A =(2k+1)\hbar.
\label{Qarea}
\eeq
Note that a $k=0$ black hole necessarily has $j=0$ and
is therefore a boson with spin zero.  

It is possible that there is
a correlation between $k$ and $j$, even away from extremality,
and that integral $j$ implies integral $k$ and half-integral $j$
implies half-integral $k$.
Indeed the area spectrum found in \cite{Gabor} for non-rotating
black holes requires integral $k$ when $j=0$ for a hole carrying
zero charge, half-integral $k$ only appear for charged black
holes in their analysis. 
The spectrum found in \cite{Medved} for zero charge requires
that $j$ and $k$ are both integral. 
The fuzzy sphere approach here does not impose any such restrictions.
While a correlation between integral $k$ and $j$, requiring that
they be either both integral or both half-integral, seems plausible
we have not found a proof that it is necessary.

The fact that the difference between the quantum bound 
(\ref{QuantumJmax}) and the classical 
bound (\ref{Jmax}) is independent of
$A$ is a direct consequence of the choice $\bar a=1$.

Equation (\ref{CRmass}) now reads
\beq
M^2
=\left\{ {k(k+1) + j(j+1) +{1\over 4}\over (2k+1)} \right\}\hbar.\eeq
The mass of a black hole of a given
area (fixed $k$) with maximum allowed angular momentum is now
\beq M^2(J_{max})={1\over 4}\left\{ 8k(k+1)+1 \over 2k+1\right\}\hbar.\eeq
In the quantum theory equation (\ref{Jmax}) is then replaced with 
\begin{equation}
J_{max}^2= M^4(J_{max})-{\hbar^4\over 16\bar A^2}
={1\over 4}(\bar A^2-\hbar^2),
\label{QuantumJbound}
\end{equation}
so (\ref{Jbound}) is never saturated for finite $k$.
In terms of $j$ and $k$ the bound is
\beq
(2k+1)^2> 4j(j+1).
\eeq

\section{Charged Black Holes}
\label{ChargedHoles}

Including electric charge $Q_e$ the classical Christodoulou-Ruffini
formula reads 
\begin{equation}
M^2={1\over \bar A}\left\{
{1\over 4} (\bar A + Q_e^2)^2 + J^2 \right\}
\end{equation}
or, if magnetic monopoles with charge $Q_m$ are also included,
\begin{equation}
M^2={1\over \bar A}\left\{
{1\over 4} (\bar A + Q^2)^2 + J^2 \right\}
\label{CRmassQ}
\end{equation}
where
\beq
Q^2=Q_e^2+Q_m^2.
\eeq
With $\bar A=(2k+1)\hbar$ and
$Q_e$ quantised in multiples of the electric
charge $e$
the quantum version of (\ref{CRmassQ}) becomes
\begin{equation}
M^2
=\left\{ {\Bigl[2k+ 1+\alpha q_e^2+ \alpha^{-1}(q_m/2)^2\Bigr]^2
+4j(j+1) \over 4(2k+1)} \right\}\hbar,
\label{Massformula}
\end{equation}
with $q_e$ and $q_m$ integers (we use units with $4\pi\epsilon_0=1$
so that the fine structure constant is $\alpha=e^2/\hbar$ when $c=1$,
the factor of $\alpha^{-1}/4$ multiplying $q^2_m$ allows for the
Dirac quantisation condition, 
$Q_eQ_m=\widetilde N\hbar/2$ where $\widetilde N$ is an integer). 
Thus, as suggested in \cite{Area}, the black hole
mass is characterised by four discrete numbers: 
$k$ and $j$, which can each be either integral or half-integral,
and $q_e$ and $q_m$ which are both integers.

This particle picture of black holes 
has also been a central theme in the work of 't~Hooft,
\cite{tHooft2} \cite{tHooft1}.
The general form of the spectrum (\ref{Massformula}) was derived by
Bekenstein \cite{Area}, the new ingredient here is that
some of the constants differ as
a consequence of the hypothesis that the event horizon is modeled
by a fuzzy sphere.

Demanding that $\bar A$ in (\ref{CRmassQ}) is real gives the
classical bound
\beq
M^4-Q^2M^2-J^2\ge 0
\label{MforJ}
\eeq

Defining
\beq\Delta^2:=M^4 - Q^2M^2 -J^2
\eeq
(\ref{CRmassQ}) can be used to express $\Delta^2$ in terms of the area
\beq
\Delta^2={(\bar A^2 - Q^4 - 4J^2)^2\over 16\bar A^2}.
\eeq
Actually for the outer horizon
\beq
\bar A^2\ge Q^4 +4J^2,
\eeq
so we can conclude that
\beq
\Delta={\bar A^2 - Q^4 - 4J^2\over 4\bar A}\ge 0.
\label{QuantumD}
\eeq
The classical bound (\ref{MforJ}) is thus saturated
when 
\beq
\bar A^2=Q^4+4J^2.
\eeq
Quantum mechanically this bound cannot always be achieved, the best
we can hope to do is minimise $\Delta$. The quantum version of $\Delta$
is, using (\ref{Qarea}),
\beq
\Delta/\hbar={(2k+1)^2 - \left(\alpha q_e^2 + \alpha^{-1}(q_m/2)^2\right)^2
-4j(j+1) \over 4(2k+1)}.
\label{QuantumBound}
\eeq
The special case $q_e=q_m=0$ reproduces the
analysis in the previous section.

The final conclusion here is that, for $q_e$, $q_m$ and $j$ given,
$k$ (or equivalently the area)
is bounded below by
\beq
(2k+1)^2 \ge \Bigl(\alpha q_e^2 + \alpha^{-1}(q_m/2)^2\Bigr)^2 + 4j(j+1).
\eeq

\section{Entropy}
\label{EntropySection}

Using Hawking's result for the entropy, at least
for large mass black holes, we have
\beq S={A\over 4\LP^2}=\pi {{\bar A}\over \LP^2} =(2k+1)\pi,
\label{Entropy}\eeq
in units with $k_B=1$.

In the fuzzy sphere picture presented here it seems natural to
guess that a $k=0$ black hole should have zero entropy,
since it does not appear to have any internal degrees of
freedom.  The simplest modification of (\ref{Entropy})
compatible with this hypothesis is to take
\beq
S=2k\pi={A\over 4\LP^2}-\pi,
\eeq
but then the number of microstates would not be an integer
in general.

In any case we expect the number of
microstates of the black hole for large $k$ to be
\beq \Omega\approx \exp(2k\pi),\eeq
using $S=\ln\Omega$.
Without a more detailed understanding of the microscopic
states however, this formula cannot be verified directly. 

String theory provides a way of calculating the entropy
of a black hole directly from the number of microscopic states.
The first calculations, \cite{Vafa},
were done for 5-dimensional black holes,
with 3-dimensional event horizons $S^3$ which are not symplectic
manifolds (fuzzy descriptions of $S^3$ do exist though, \cite{Ramgoolam}). A string theory derivation of the entropy of
extremal supersymmetric black holes in 4-dimensions
was given in \cite{MalStrom} and generalised to the
non-extremal case in \cite{Horowitz}.
The entropy calculated in \cite{MalStrom} 
agreed with the earlier explicit evaluations of the black hole
area \cite{BPZhole} and reproduced Hawking's factor $1/4$.
The upshot of the analysis in \cite{MalStrom}
is that the entropy depends on four integers,
labelled $Q_2$, $Q_6$, $n$ and $m$ in their notation,
and, for large integers, is given by
\beq
S=2\pi\sqrt{Q_1Q_6nm}.
\eeq
Clearly this agrees with the result (\ref{Entropy}) if, at least for
large $k$,
\beq
k^2\approx Q_1Q_6nm.
\eeq
In general one expects corrections to this formula of order $k$.

It is not obvious how the string theory arguments 
might relate to (\ref{AreaN}).  
The problem is that the string theory calculation must be 
carried out in the regime of small string coupling $g_s$, where
the notion of a black hole is not well defined.  
Black holes, it is believed,
emerge from string theory as classical objects only for large $g_s$ 
and one relies on supersymmetry to argue that the small $g_s$ 
calculation still gives the correct answer for the entropy
even when $g_s$ is large.
But there is no analogue of (\ref{AreaN}) in string theory when
$g_s$ is small.
It has been suggested that fuzzy spheres can be viewed as spherical
D2-branes \cite{Dbranes} and they also emerge as
ground states of matrix models \cite{MatrixModels}, so
it may prove possible to investigate the ideas presented here
directly in string theory.  

Attempts have also been made to calculate the entropy of black holes in
the loop approach to quantum gravity 
(see \cite{LoopGravity} and references
therein).
An equal spaced area spectrum like that of (\ref{FourPiAreaLaw}) 
was found in
\cite{Polychronakos} though the prefactor was undetermined,
and the question of
sub-leading corrections 
has also been addressed \cite{Meissner}.
For a discussion of entropy from a fuzzy sphere approach within
the loop quantum gravity framework see \cite{Paola}.

\section{Conclusions}
\label{Conclusions}

By modeling the event horizon of a black hole as a fuzzy sphere
and assuming an equally spaced area law $A =N l_P^2$ 
it has been shown that,
in the continuum limit as $N\rightarrow\infty$,
the event horizon looks locally like
a non-commutative plane with non-commutativity parameter
\beq \theta={a \LP^2\over 4\pi}\eeq
independent of the mass.

If an equally spaced area law is assumed to hold even at finite $N$
a mass spectrum of a quantum black hole can be derived
by identifying the maximum angular momentum
of the black hole with the maximum angular momentum associated
with the Hilbert space underlying the fuzzy sphere, which requires
$a=4\pi$.  The spectrum is then
\begin{equation}
M_{k,j,q_e,q_m}^2
=\left\{ {\Bigl[2k+1+ \alpha q_e^2 + \alpha^{-1} (q_m/2)^2\Bigr]^2
+4j(j+1) \over 4(2k+1)} \right\}\MP^2,
\label{fullmassformula}
\end{equation}
where $k$ is either an integral or a half-integral quantum
number determining the area of the event horizon,
\begin{equation}
A=4\pi(2k+1)\LP^2,
\label{AreaQ}
\end{equation}
$j$ is the angular momentum quantum number, $q_e$ the electric charge
and $q_m$ the monopole charge.
For given values of $j$, $q_e$ and $q_m$ the
quantum number $k$ is bounded below by the quantum analogue
of the familiar classical bound,
\beq
(2k+1)^2\ge 4j(j+1) + \left(\alpha q_e^2 + \alpha^{-1} (q_m/2)^2\right)^2.
\label{kbound}
\eeq
In general the quantum bound (\ref{kbound}) cannot be saturated
unless $\alpha$ takes on special values, for example when $j=q_m=0$
the bound can be saturated if $\alpha$ is rational.
The area quantisation (\ref{AreaQ}) has also been found in
a semi-classical approach \cite{PadPat} and in
a mini-superspace approach to quantising black holes \cite{Gabor}
\cite{Medved},
but in \cite{Gabor} it is argued that $\alpha$ must be rational,
whereas the fuzzy sphere approach presented here does not seem to
require this.

Equation (\ref{fullmassformula}) 
is a version of a suggestion of Bekenstein's,
but with different constants.
The constants have been fixed here by assuming that the maximum
angular momentum of a rotating hole be identified with the
maximum angular momentum of the underlying Hilbert space
of the fuzzy sphere, thus side-stepping the question of the
microscopic degrees of freedom.  
If this assumption is relaxed,
then the above formulae still apply but with the undetermined
parameter $\bar a$ re-introduced so that $2k+1$ is replaced with
$(2k+1)\bar a$ everywhere. Bekenstein used $\bar a=(\ln 2)/\pi$ in
\cite{Area} and Hod found $\bar a=(\ln 3)/\pi$ in \cite{Hod}.  Bekenstein's
value of $\ln 2/\pi$ is a consequence of associating classical bits with
each unit of phase space area and calculating the number of possible
configurations.  In a fully quantum approach it would seem more natural to
use qubits rather than classical bits, as suggested in \cite{Paola}, and to
determine the entropy from
a density matrix, but any explicit expression would require making further, more specific,
assumptions about the allowed quantum states.  
Alternatively it may be possible to determine the entropy using a specific field theory 
on the fuzzy sphere, perhaps a supersymmetric field theory, or by viewing the fuzzy sphere as a dynamical object in the context of  a matrix model.  From this last point of view it is intriguing that studies of the
ground state fuzzy sphere in matrix Chern-Simons theory \cite{ANN}
reveal a radius that scales as $\sqrt{N}$, which is precisely what is needed for
the black-hole interpretation advanced here.

Assuming that $\bar a=1$  we see that the smallest
mass (the ground state) given by this formula is
\beq M={1\over 2}\MP=6.10\times 10^{18}\ GeV/c^2,\eeq
when $k=j=q_e=q_m=0$. 
The event horizon area for the minimum mass black hole is
\beq A=4\pi\LP^2.\eeq

The next smallest mass is for a non-rotating black hole carrying
a single unit of charge $q_e=1$ with $k=q_m=0$, which lies
\beq
\Delta M={1\over 2}\alpha \MP 
\eeq
above the ground state.  The numerical value here depends on the value
of $\alpha$ used.
One should take into the account running of the coupling constant
and use $U(1)$ hypercharge rather
than electric charge, or some other $U(1)$ charge depending on new
physics.

Our analysis has avoided any discussion of the microscopic
degrees of freedom of the black hole.  In particular little
has been said about entropy beyond using Hawking's
formula to determine the entropy from the area.  This formula
may well be modified for small black holes by quantum phenomena,
but without a more detailed understanding of the black hole microstates
it is impossible to be more specific at this stage.

This work was partly funded by an EU Research Training Network grant
in Quantum Spaces-Noncommutative geometry QSNQ, and partly
by an Enterprise Ireland Basic Research grant SC/2003/415.


\begin{thebibliography}{04}

\bibitem{Bekenstein} J.D.~Bekenstein, {\it Lett.~Nuovo.~Cimento} 
{\bf 4} (1972) 737;\hfill\break
J.D.~Bekenstein, {\it Phys. Rev.} {\bf D7} (1973) 2333 

\bibitem{Hawking} S.W.~Hawking, {\it Nature} {\bf 248} (1974) 30;\hfill\break
S.W.~Hawking, {\it Comm.~Math.~Phys.} {\bf 43} (1975) 199;\hfill\break 
S.W.~Hawking, {\it Phys. Rev.} {\bf D13} (1976) 191


\bibitem{Sorkin} R.D.~Sorkin, {\it 
On the Entropy of the Vacuum Outside a Horizon},
in  {\it Tenth International Conference on General Relativity and Gravitation
(Padova, 4-9 July, 1983), Contributed Papers}, 
vol. II, eds. B. Bertotti, F. de Felice and A. Pascolini, (1983) 734
(Roma, Consiglio Nazionale Delle Ricerche);
%

\bibitem{Madore} J.~Madore, {\it Class.~Quant.~Grav.} {\bf 9} (1992) 69

\bibitem{Particle} J.D.~Bekenstein,
{\sl Quantum Black Holes as Atoms},
Proceedings of the 8th Marcel Grossmann Meeting,
ed.~T.~Piran, World Scientific (1988),
{\tt gr-qc/9710076}

\bibitem{Area} J.D.~Bekenstein, {\it Lett.~Nuovo~Cimento} {\bf 11} (1974) 467; {\sl Black Holes: Classical Properties, Thermodynamics
and Heuristic Quantisation}, {\tt gr-qc/9808028}

\bibitem{tHooft2} G.~'t~Hooft, {\sl The Holographic Principle},
{\tt hep-th/0003004}; \hfill\break
G.~'t~Hooft, {\it Class.~Quant.~Grav.} {\bf 13} (1996) 1023,
{\tt gr-qc/9601014} 

\bibitem {tHooftBombelli} G.~'tHooft,  {\it Nuc. Phys.} {\bf B256} (1985) 727;
L.~Bombelli, Rabinder K.~Koul, Joohan Lee and R.D.~Sorkin, 
{\it Phys. Rev.~D} {\bf 34} (1986) 373
%

\bibitem{OtherArea}V.~Mukhanov, {\it JETP Lett.} {\bf 44} (1986) 63; 
I.I.~Kogan, {\it JETP~Lett.} {\bf 44} (1986) 267, {\tt hep-th/9412232};
M.~Maggiore, {\it Nucl.~Phys.~B} {\bf 429} (1994) 205;
J.D.~Bekenstein and V.F.~Mukhanov,
{\it Phys.~Lett.} {\bf B360} (1995) 7

\bibitem{tHooft1} G.~`t~Hooft, {\sl The Black Hole 25 years after}, 
{\tt gr-qc/9402037}; \hfill\break
G.~'t~Hooft, {\it Int.~J.~Mod.~Phys.} {\bf A} (1996) 4623 

\bibitem{MadoreBook} J.~Madore, {\sl An Introduction to Non-commutative 
Differential Geometry and its Physical Applications},
2nd edition, Cambridge University Press (1999)

\bibitem{PadPat} T.~Padmanabhan and A.~Patel, {\sl Semi-classical
quantization of gravity I: Entropy of horizons and the area
spectrum.}, {\tt hep-th/0305165}

\bibitem{Gabor} A.~Barvinsky, S.~Das and G.~ Kunstatter, 
{\it Class.~Quant.~Grav.\/} {\bf 18} (2001) 4845, {\tt gr-qc/0012066}; 
{\it Phys.~Lett.\/} {\bf B517} (2001) 415, {\tt hep-th/0102061}; 
{\it Found.~Phys.\/} {\bf 32} (2002) 1851, {\tt hep-th/0209039}; 
S.~Das, P.~Ramadevi and U.A.~Yajnik, {\it Mod.~Phys.~Lett.\/} 
{\bf A17} (2002) 993, {\tt hep-th/0202076}; 
S.~Das, P.~Ramadevi, U.A.~Yajnik and A.~Sule, {\it Phys.~Lett.\/} 
{\bf B565} (2003) 201, {\tt hep-th/0207169}; 
S.~Das, H.~Mukhopadhyay and P.~Ramadevi, 
{\sl Spectrum of rotating black holes and its implications 
for Hawking radiation\/}, {\tt hep-th/0407151}

\bibitem{Medved} G.~Gour and A.J.M.~Medved, 
{\it Class.~Quant.~Grav.} {\bf 20} (2003) 1661, {\tt gr-qc/0212021};
{\it ibid.\/} 
{\bf 20} (2003) 2261, {\tt gr-qc/0211089}

\bibitem{Polychronakos} A.~Alekseev, A.P.~Polychronakos 
and M.~Smedb\"{a}ck, {\it Phys.~Lett.} {\bf B574} (2003) 296,
{\tt hep-th/0004036}

\bibitem{Romero} J.M.~Romero, J.A.~Santiago and J.D.~Vergara, 
{\it Phys.~Rev.} {\bf D68} (2003) 067503,
{\tt hep-th/0305080}

\bibitem{Li} Miao~Li, {\tt hep-th/0006024};
Miao~Li, {\it Class.~Quant.~Grav.} {\bf 21} (2004) 3571,
{\tt hep-th/0311105};
A.~Krause, {\sl On the Bekenstein-Hawking Entropy,
Non-commutative Branes and Logarithmic Corrections}, 
{\tt hep-th/0312309} 

\bibitem{BuricMadore} M.~Buri\'c and J.~Madore, 
{\it Noncommutative 2-Dimensional Models of Gravity}, {\tt hep-th/0406232}

\bibitem{Perelomov} A.~M.~Perelomov,
``Generalized Coherent States and their Applications'', Springer (1986)

\bibitem{Al} G.~Alexanian, A.~Pinul and A.~Stern, {\it Nucl.~Phys.} {\bf B600} (2001) 531,
{\tt hep-th/0010187}

\bibitem{Chuetal} Chong-Sun Chu, J.~Madore and H.~Steinacker,
{\it JHEP} {\bf 0108} (2001) 038, {\tt hep-th/0106205}

\bibitem{Christodoulou} D.~Christodoulou and R.~Ruffini, {\it Phys.~Rev.}
{\bf D4} (1971) 3552
 
\bibitem{Vafa} A.~Strominger and C.~Vafa, 
{\it Phys.~Lett.} {\bf B379} (1996) 99 {\tt hep-th/9601029};
C.G.~Callan and J.M.~Maldacena, {\it Nucl.~Phys.} {\bf B472} (1996) 591,
{\tt hep-th/9604023}

\bibitem{Ramgoolam} S.~Ramgoolam, {\it Nucl.~Phys.}~{\bf B610}~(2001)~461,
{\tt hep-th/0105006};
B.P.~Dolan, P.~Pre\v{s}najder and D.~O'Connor, 
{\it JHEP} {\bf 0402} (2004) 055, {\tt hep-th/0312190}

\bibitem{MalStrom}
J.M.~Maldacena and A.~Strominger,  
{\it Phys.Rev.Lett.} {\bf 77} (1996) 428, {\tt hep-th/9603060}

\bibitem{Horowitz} G.T.~Horowitz, D.A.~Lowe and J.M.~Maldacena,
{\it Phys.~Rev.~Lett.} {\bf 77} (1996) 430, {\tt hep-th/9603195}

\bibitem{BPZhole} R.~Kallosh, A.~Linde, T.~Ortin, A.~Peet and A.~van~Proeyen,
{\it Phys.~Rev.} {\bf D46} (1992) 5278;
M.~Cvetic and D.~Youm, {\it Phys.~Rev.} {\bf D53} (1996) 584, 
{\tt hep-th/9507090};
M.~Cvetic and A.~Tseytlin, {\it Phys.~Rev.} {\bf D53} (1996) 5619, 
{\tt hep-th/9512031};
erratum {\it ibid.} {\bf D55} (1997) 3907;
R.~Kallosh and B.~Koll, {\it Phys.~Rev.} {\bf D53} (1996) 5344, 
{\tt hep-th/9602014}

\bibitem{Dbranes} A.Yu.~Alekseev, A.~Recknagel and V.~Schomerus,
{\it JHEP} {\bf 9909} (1999) 023, {\tt hep-th/9908040};
Y.~Hikida, M.~Nozaki and Y.~Sugawara, 
{\it Nucl.Phys.} {\bf B617} (2001) 117, 
{\tt hep-th/0101211}; 
K.~Hashimoto and K.~Krasnov, {\it Phys.Rev.} {\bf D64} (2001) 046007, 
{\tt hep-th/0101145}

\bibitem{MatrixModels} S.~Iso, Y.~Kimura, K.~Tanaka and
K.~Wakatsuki, {\it Nucl.~Phys.} {\bf B604} (2001) 121, 
{\tt hep-th/0101102} 

\bibitem{LoopGravity} L.~Smolin, {\it J.~Math.~Phys.} {\bf 36} (1995) 6417;\hfill\break
C.~Rovelli, {\it Phys.~Rev.~Lett.} {\bf 77} (1996) 3288;\hfill\break
K.~Krasnov, {\it Phys.~Rev.} {\bf D55} (1997) 3505;\hfill\break
A.~Ashtekar, J.~Baez, A.~Corichi, K.~Krasnov, {\it Phys.~Rev.~Lett.} {\bf 80} (1998) 904;
A.~Ashtekar, J.~Baez, K.~Krasnov, {\it Adv.~Theor.~Math.~Phys.} {\bf 4} (2000) 1

\bibitem{Meissner} K.A.~Meissner,
{\sl Black Hole Entropy in Loop Quantum Gravity},
{\tt gr-qc/0407052}

\bibitem{Paola} P.A.~Zizzi,
{\sl A Minimal Model for Quantum Gravity},
{\tt gr-qc/0409069}

\bibitem{Hod} S.~Hod, {\it Phys.~Rev.~Lett.} {\bf 81} (1998) 4293, {\tt gr-qc/9812002}


\bibitem{ANN} T.~Azuma, K.~Nagao and J.~Nishimura, {\it Perturbative Dynamics of Fuzzy Spheres at large N}, {\tt hep-th/0410263}; 
P.~Castro-Villarreal, R.~Delgadillo-Blando and B.~Ydri, 
{\it Nuc. Phys.} {\bf B704} (2005) 111, {\tt hep-th/0405201}

\end{thebibliography}
\end{document}